\documentclass[aps, prd, preprint, superscriptaddress, tightenlines, nofootinbib]{revtex4}

\pdfoutput=1

\usepackage{amsmath}	
\usepackage{amsthm}	
\usepackage{amssymb}	
\usepackage{datetime}
\usepackage{graphicx}
\usepackage{color}
\usepackage{verbatim}

\definecolor{grey}{rgb}{0.4,0.4,0.4}
\definecolor{dullmagenta}{rgb}{0.4,0,0.4}
\definecolor{darkblue}{rgb}{0,0,0.4}
\definecolor{midblue}{rgb}{0,0,0.5}
\definecolor{midred}{rgb}{0.5,0,0}
\definecolor{orange}{rgb}{1,0.5,0}
\definecolor{lightbrown}{rgb}{0.75,0.5,0.25}
\definecolor{tan}{cmyk}{0.14,0.42,0.56,0}
\definecolor{djunglegreen}{cmyk}{0.99,0,0.52,0}
\definecolor{lightgreen}{rgb}{0,1,0}
\definecolor{olivegreen}{cmyk}{0.64,0,0.95,0.40}
\definecolor{midgreen}{rgb}{0.0,0.675,0.0}
\definecolor{darkgreen}{rgb}{0,0.5,0}

\newcommand{\vs}{\vspace}

\renewcommand{\.}{\hspace{0.5mm}}

\newcommand{\Vrm}{\ensuremath{\mathrm{V}}}

\newcommand{\Ocal}{\ensuremath{\mathcal{O}}}

\newcommand{\defas}{\mathrel{\mathop :}=} 

\renewcommand{\d}{\ensuremath{\mathrm{d}}}

\newcommand{\eg}{e.g.}
\newcommand{\ie}{i.e.}

\newcommand{\cf}{c.f.}

\let\baraccent=\= 
\renewcommand{\=}[1]{\stackrel{#1}{=}} 

\theoremstyle{definition}

\theoremstyle{remark}

\settimeformat{ampmtime}

\usepackage{float}

\begin{document}

\title{Corpuscular Consideration of Eternal Inflation}

\author{Florian K{\"u}hnel}
\email{florian.kuhnel@fysik.su.se}
\affiliation{The Oskar Klein Centre for Cosmoparticle Physics,
	Department of Physics,
	Stockholm University,
	AlbaNova,
	106\.91 Stockholm,
	Sweden}

\author{Marit Sandstad}
\email{marit.sandstad@astro.uio.no}
\affiliation{Institute of Theoretical Astrophysics,
	University of Oslo,
	P.O.~Box 1029 Blindern,
	N-0315 Oslo,
	Norway}

\date{\formatdate{\day}{\month}{\year}, \currenttime}

\begin{abstract}
We review the paradigm of eternal inflation in the light of the recently proposed corpuscular picture of space-time. Comparing the strength of the average fluctuation of the field up its potential with that of quantum depletion, we show that the latter can be dominant. We then study the full respective distributions in order to show that the fraction of the space-time which has an increasing potential is {\it always} below the eternal-inflation threshold. We prove that for monomial potentials eternal inflaton is excluded. This is likely to hold for other models as well.
\end{abstract}

\maketitle

\section{Introduction}
\label{sec:Introduction}
\setcounter{equation}{0}

\noindent Cosmological inflation \cite{Starobinski,Guth:1980zm} is one of the central building blocks of our current understanding of the Universe. One of its simplest realizations, which is still compatible with observations, is via a single scalar field, called inflaton. Today's structure in the Universe is seeded by the quantum fluctuations of this field and of the space-time, and is in remarkable agreement with measurements (\cf~\cite{Planck}). Depending on the value of the inflaton, it might experience large quantum fluctuations, also and in particular, up its potential, therefore inducing ever expanding inflationary patches of the Universe. This is the idea behind eternal inflation \cite{Vilenkin:1983xq, Linde:1986fd} (\cf~for a recent review \cite{Guth:2007ng}).

These considerations are, however, only valid if the semi-classical description of space-time is a faithful approximation. If gravity, like {\it all} other fundamental interactions, possesses a quantum description, inevitably the question arises when such a corpuscular picture of space-time starts to become relevant. Recent progress by Dvali and Gomez (\cf~\cite{Dvali:2013eja, DvaliGomez, Dvali:2014gua}), elaborating precisely on this topic, suggests that in certain situations one necessarily needs to take the graviton nature of what is classically regarded as space-time geometry into account. In fact, some phenomena like Hawking radiation, the Bekenstein Entropy, or, the information paradox can only be fully understood in this quantum picture \cite{Dvali:2013eja, DvaliGomez, Dvali:2014gua} (\cf~also \cite{Flassig:2012re, Casadio:2015xva, Kuhnel:2014oja, mueckPT, Hofmann, Binetruy:2012kx, Kuhnel:2014gja, Casadio:2014vja} for recent progress).

The mentioned attempt leads to regard space-time, such as black holes, de Sitter spaces, etc., as gravitationally bound states in the form of weakly/marginally bound states, or Bose-Einstein condensates, of gravitons with a mean wave-length equal to the curvature radius of that space-time. Due to the weak binding, quantum fluctuations are responsible for emptying the ground state of the condensate. This depletion is an intrinsically quantum effect which is entirely missed in any (semi-)classical treatment. In inflationary spaces, it acts like a quantum clock which works against the semi-classical one and, as we will investigate below, also against the fluctuations of the scalar field up the potential.

Recently, in \cite{Dvali:2014gua} it has been argued that the corpuscular picture is incompatible with a positive cosmological constant. Also, in Ref.~\cite{Dvali:2013eja} the authors argue that quantum depletion sets a limit on the total number of e-foldings. Here, we investigate those qualitative considerations in more detail by quantitatively comparing the strengths of the relevant effects and considering, via the full respective probability distributions, the fraction of the space-time which has an increasing potential.

The mentioned corpuscular picture of quantum gravity, as introduced in \cite{Dvali:2013eja} (\cf~also \cite{DvaliGomez, Dvali:2014gua}), assumes gravitons on a Minkowski background. Since they are bosons, and given their peculiar attractive derivative self-coupling, they generically form Bose-Einstein condensates. In the limit of very high graviton ground-state occupation number $N$, these condensates yield the emergent geometry which is observed at the classical level.

Certain of these Bose-Einstein condensates are very special as their particular densities put them at a point of quantum criticality. This criticality occurs when the interaction strength $\alpha$ is inversely proportional to the number of gravitons present in the condensate. In turn, these states can be defined in terms of a scale, which is also the characteristic length scale of the system. The self-interactions of the gravitons make this critical phase stable. This means that removal or addition of gravitons to the critical condensate happens self-similarly{\,--\,}changing the defining length scale of the system while remaining at criticality. The defining length scale is proportional to the square root of the number of gravitons in the critical condensate. A prime example of such a state is a Schwarzschild black hole, where the defining length scale is the Schwarzschild radius $R_{S}$ and the graviton number is $N = ( R_{S} / L_{P} )^{2}$.

The de Sitter or inflationary patch is another such condensate at criticality, where the defining length scale is the Hubble radius $R_{H}$ and the graviton number $N = ( R_{H} / L_{P} )^{2}$. This inflationary case is somewhat more complicated than the black hole case, as the inflationary state is a composite, comprising of a critical graviton Bose-Einstein condensate interacting with a much higher-occupied inflaton condensate.

Though in the large-$N$ limit, the emergent geometries are classical, the above outlined corpuscular description is fully quantum at heart. Belonging to the quantum critical condensate, the ground state of the gravitons has nearby, tightly-spaced Bogoliubov states, accessible through graviton-to-graviton or graviton-to-inflaton scattering. In the inflationary case, inflatons vastly outnumber gravitons, but inflaton self-interactions cannot excite the inflatons from the ground state, and hence graviton-to-inflaton scattering is the dominant quantum processes which deteriorates the classical geometry, and thereby also the anchor point of all semi-classical computations.

In \cite{Dvali:2013eja} it was argued qualitatively that this quantum depletion of the condensates excludes eternal inflation. Here we will expand on that argument to consider eternal inflation for this corpuscular description of quantum gravity to do a quantitative exploration of the subject. We include all monomial potentials, not only the $m^{2}\.\phi^2$-version that was considered in \cite{Dvali:2013eja}.

\section{Competing Fluctuations}
\label{sec:Main-Part}
\setcounter{equation}{0}

\noindent We consider a Universe filled with inflaton and graviton condensates. The number of coherent inflatons in the inflaton condensate is $N_\phi$ and the number of coherent gravitons in the graviton condensate is $N$.\footnote{We will sometimes refer to the number of coherent inflatons/gravitons in the condensates only as the number of inflatons/gravitons, as this is the only numbers of particles we will be interested in.} Working in Planck units ($c = \hslash = M_{\rm Pl} = 1$), the number of coherent inflatons can easily be defined as
\begin{equation}
	N_{\phi}
		\defas
								n_{\phi}\.R_{H}^3
		=
								\frac{n_{\phi} }{ H^3}
								\, ,
\end{equation}
where $n_{\phi}$ is the number density of inflatons in the condensate. The number of coherent gravitons is given by
\vs{-2mm}
\begin{equation}
	N
		=
								R_{H}^{2}
		=
								\frac{1}{H^{2}}
								\; .
\end{equation}

When considering eternal inflation in view of the corpuscular description of gravity, we find that two competing quantum effects are active:~the quantum fluctuations of the inflaton field due to the uncertainty principle, and the quantum depletion of the inflaton and graviton condensates due to graviton-inflaton scattering.

The typical quantum fluctuation due to the uncertainty principle reads (\cf~\cite{Vilenkin:1983xq})
\begin{equation}
	|\dot{\phi}|
		=
								\frac{H^2}{2\pi}
								\, .
								\label{eq:qfluct}
\end{equation}
With interaction strength of $\alpha=\frac{1}{N}$, combinatoric factors of $NN_\phi$ and $N(N-1)$ for graviton-inflaton and graviton-graviton scattering respectively, the quantum depletion of the coherent gravitons is, to leading order in $1 / N$, (\cf~\cite{Dvali:2013eja})
\begin{align}
	\dot{N}
		&\simeq
								-
								\frac{1}{\sqrt{N\.}}\.\frac{N_\phi}{N}
								-
								\frac{1}{\sqrt{N\.}}\.N_{\mathrm{ls}}
								\, .
								\label{eq:GravitonDep}
\end{align}
Here $N_{\mathrm{ls}}$ is the number of species that are lighter than the energy of the gravitons in the condensate, which then present a possible decay channel. If no such lighter species exist the second term in Eq.~\eqref{eq:GravitonDep} will just be $- 1 / \sqrt{N\.}$ and represents graviton-graviton scattering. 

The quantum fluctuations in the inflaton medium is the only source that may increase the energy of the inflaton medium, pushing the inflaton upwards in the potential. For eternal inflation to be realised, this effect has to be larger than the effect of depletion in a large enough fraction of the space, so that combined with the continued inflation of this fraction of space it increases the volume of the inflating part of space-time. Since for each Hubble time this increase in volume is $e^{3}$, this fraction must be $< e^{- 3} \approx 1/ 20 = 0.05$.

To compare the two effects we look for simplicity at an inflaton in the monomial potential
\begin{equation}
	\Vrm( \phi )
		=
								\frac{1}{n!}\.\lambda_{n}\.\phi^{n}
								\, ,
								\label{eq:V}
\end{equation}
where $n > 0$, and we define the effective mass scale
\begin{align}
	m_{\rm eff}( \phi )
		&\equiv
								\sqrt{\big| \Vrm''( \phi ) \big|\,}
								\; ,
								\label{eq:meff}
\end{align}
which for the case of $\Vrm( \phi ) = 1 / 2\.m^{2}\.\phi^{2}$ yields $m_{\rm eff}( \phi ) \equiv m_{\rm eff} = m$. We shall work in the slow-roll regime, $|\dot\phi^{2}| \ll | \Vrm |$ and $| \ddot\phi^{2} | \ll 3\.H\.\dot\phi \sim | \Vrm_{\!,\phi} |$, and use Eq.~\eqref{eq:GravitonDep}, the Friedmann equation $H^{2} = 1 / 3\.\Vrm$, as well as $N_{\phi} \simeq ( \Vrm / m_{\rm eff} )\.R_{H}^{3}$, to obtain the useful relations
\begin{subequations}
\begin{align}
	\phi
		&\simeq
								\sqrt[n\,]{\frac{3\.n!}{N\.\lambda_{n}}\.}
								\; ,
								\displaybreak[1]
								\\[2.5mm]
	N_{\phi}
		&\simeq
								\sqrt[n\,]{\frac{ 3\.n! }{ N\.\lambda_{n} }\,} \sqrt{\frac{ 3 }{ n | n - 1 | }\,}\,N
								\; ,
								\label{eq:NPhiOfN}
								\displaybreak[1]
								\\[1.5mm]
	\dot{\phi}
		&\simeq
								- \frac{ 1 }{ n }\.\sqrt[n\,]{\frac{3\.n!}{N\.\lambda_{n}}\.}\,\frac{ \dot{N} }{ N }
								\; ,
								\label{eq:dotPhiofDotN}
								\displaybreak[1]
								\\[2.5mm]
	m_{\rm eff}
		&\simeq
								\sqrt[n\,]{\frac{N\.\lambda_{n}}{3\.n!}\.}\.\sqrt{\frac{ 3\.n | n - 1 | }{ N }\,}
								\; ,
								\label{eq:meffOfN}
\end{align}
\end{subequations}
where we assume that $n \ne 1$.

For quantum fluctuations in Eq.~\eqref{eq:qfluct} to move the inflaton up the potential, the fluctuations have to be larger than the average depletion \eqref{eq:GravitonDep} which deteriorates the state away from semi-classicality. Using Eq.~\eqref{eq:NPhiOfN} for $N_{\phi}$ in Eq.~\eqref{eq:GravitonDep}, we realise that the magnitude of depletion is:
\begin{equation}
	| \dot{N}_{\rm dep} |
		\simeq						
								\sqrt[n\,]{\frac{ 3\.n! }{ N\.\lambda_{n} }\,} \sqrt{\frac{ 3 }{ n | n - 1 |\.N }\,}
		=	
								\frac{ 3 }{ m_{\rm eff}\.N }\,
								\, .
								\label{eq:Ndot<3/m}
\end{equation}

We then insert Eqs.~\eqref{eq:dotPhiofDotN} and \eqref{eq:meffOfN} into Eq.~\eqref{eq:qfluct} to find the contribution to $\dot{N}$ stemming from a typical quantum fluctuation:
\begin{equation}
	|\dot{N}_{\rm qf}|
		\simeq
								\frac{ n }{ 2 \pi }\.\sqrt[n\,]{\frac{ N\.\lambda_{n} }{ 3\.n! }\,}
		=
								\frac{ 1 }{ 2 \pi }\.\sqrt{\frac{ n }{ 3\.| n - 1 | }\,}\;m_{\rm eff}\.\sqrt{N}
								\; .
								\label{eq:Ndot=Nm/sqrt6pi}
\end{equation}
For eternal inflation to make any sense, the quantum-fluctuating effects that drive it must be dominating the effects of the quantum depletion of the condensates. The two effects might seem, in the case when the quantum fluctuations drive the inflaton up its potential, to both push the condensate towards lower values of $N$. However, the way in which this happens is different. The depletion drives the condensate away from its classical inflating description towards an intrinsically quantum state. For this the coherent graviton-condensate behaviour which constitutes the inflating geometry no longer exists, and the description of this as inflation (eternal or otherwise) disappears. Comparing \eqref{eq:Ndot<3/m} and \eqref{eq:Ndot=Nm/sqrt6pi} we find that the condition for the quantum fluctuations being dominant reads
\begin{align}
	N^{3/4} m_{\rm eff}
		&\gtrsim
								\sqrt{ 6 \pi\.\sqrt{\frac{ 3 | n - 1 |}{ n }\,}\,}
								\; .
								\label{eq:N>1/m}
\end{align}

In order for eternal inflation to proceed when no corpuscular effects are present, the standard deviation of the Gaussian-distributed quantum fluctuations must be large enough for $1 / e^{3} \approx 1 / 20$ of the fluctuations to exceed the classical roll down the potential. This gives the criterion: $H^{2} / | \dot{\phi}_{\rm cl} | \gtrsim 3.8$ (see \cite{Guth:2007ng}). Inserting for the corpuscular variables and our potential \eqref{eq:V}, this translates to the demand:
\vspace{-3mm}
\begin{align}
	N\.m_{\rm eff}
		&\lesssim
								\sqrt{\frac{3| n - 1 |\.}{n}}\frac{1}{3.8}
								\; .
								\label{eq:N<1/m}
\end{align}
To have eternal inflation, this bound must be fulfilled while the quantum-depletion effects are still smaller than the regular quantum fluctuations. Taking the condition given for the vacuum fluctuations to dominate over depletion \eqref{eq:N>1/m} and demanding that it holds above the bound yielding eternal inflation classically \eqref{eq:N<1/m}, we find
\begin{equation}
	\frac{| n - 1 | }{ 2500\.\pi^{2} n}
		\gtrsim
								N
								\; .
\end{equation}
Since $N$ can never go below $1$, na{\"\i}vely, eternal inflation can never take place, because regardless of the value of $n$, this bound will always imply that $N$ is much smaller than $1$, unless $n$ is extremely close to zero, which means that the potential is extremely flat. In fact, formally
\begin{equation}
	n
		\lesssim
								4\times10^{-5}
								\; .
								\label{eq:naive-bound-on-n}
\end{equation}
in order for eternal inflation to have a chance of dominating before depletion takes over, hence, na{\"\i}vely we could already exclude the occurrence of eternal inflation for all monomial potentials. 



Irrespective of the value of $n$, the above calculation was just done by comparing the {\it typical} quantum fluctuation and depletion. In order to get a more refined exclusion of eternal inflation, we must consider the two distributions properly. The quantum fluctuations approximately follow a Gaussian distribution centered around zero. The depletion process is given by a Poisson distribution \cite{Kuhnel:2014oja}.

For safe bounds on eternal inflation, we can compare the two distributions at the lower bound for the potential, so at the value of $N\.m_{\rm eff}$ where standard (non-corpuscular) eternal inflation would occur [\cf~Eq.~\eqref{eq:N<1/m}]. Inserting this into Eq.~\eqref{eq:Ndot=Nm/sqrt6pi} we find that here the Gaussian distribution has a standard deviation $\sigma \simeq (n/2\pi)(3.8n)^{-2/(2+n)} (\lambda_n/3n!)^{1/(2+n)}$, whereas Eq.~\eqref{eq:Ndot<3/m} implies that the Poisson distribution has an expectation value of $\lambda \simeq 3.8\.\sqrt{3 n/ | n - 1 | \,} $.

\begin{figure}
	\centering
	\vs{-2mm}
	\includegraphics[scale=1.2,angle=0]{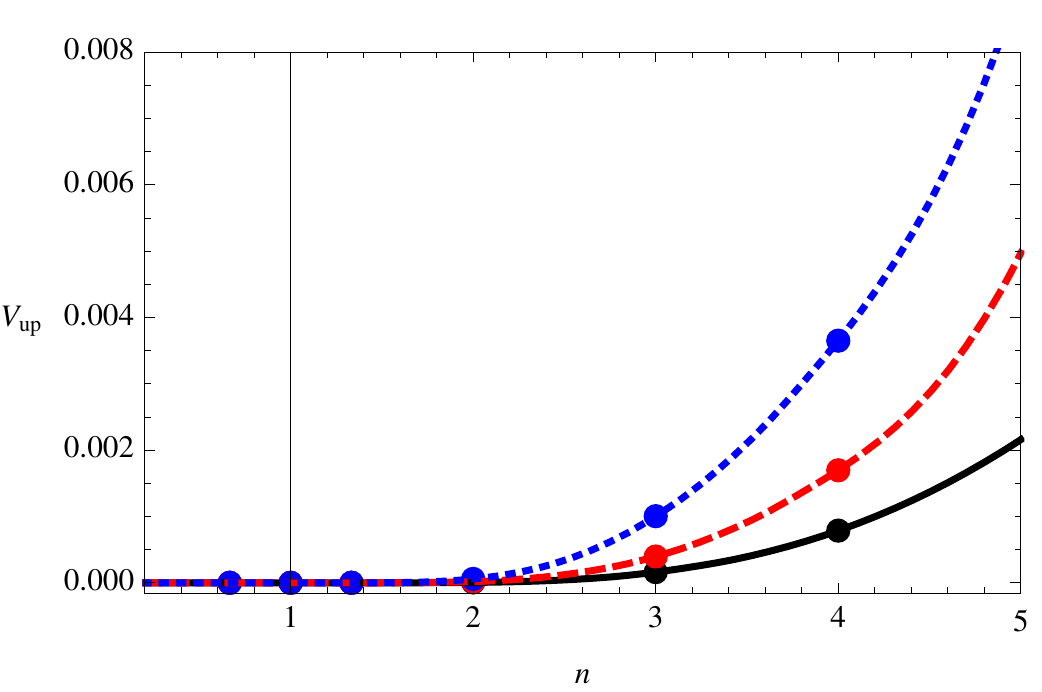}
	\caption{The convolution integral $\Vrm_{\mathrm{up}}$ [\cf~Eq.~\eqref{eq:Vup}]
			as a function of the exponent $n$ in $\Vrm( \phi ) = 1 / n!\.\lambda_{n}\.\phi^{n}$,
			for various values of the self-coupling $\lambda_{n}$:
			$10^{-8}$ (blue, dotted), $10^{-10}$ (red, dashed), $10^{-12}$ (black, solid).
			}
	\label{fig:Vup}
\end{figure}

In order to compare the two competing effects we must then convolute the two probability distribution functions to find the fractional convoluted area that gives an increase in inflaton energy. That is, for each possible value for depletion, we multiply its probability with the probability of all quantum fluctuations that are large enough to dominate over it. In practice this is done by integrating the Gaussian up to where its contribution is the negative of each point on the Poisson curve along the Poisson distribution:
\begin{align}
	\Vrm_{\mathrm{up}}
		& \simeq
								\int_{0}^{\infty}\d t\;
								\frac{ \lambda^{t}\,e^{-\lambda} }{ t! }
								\left[
									\int_{-\infty}^{-t}\d x\;
									\frac{1}{\sigma\sqrt{2\pi}} e^{- \frac{x^{2}}{2 \sigma^{2}}}
								\right]
								.\label{eq:Vup}
\end{align}
$\Vrm_{\rm up}$ is the fraction of the space-time that has an increasing potential. Since the space-time volume in the inflating parts of the Universe is multiplied by twenty, eternal inflation can only occur when $\Vrm_{\mathrm{up}} \gtrsim 0.05$. 
Fig.~\ref{fig:Vup} shows that, at least for small $n$, eternal inflation is excluded. 
Note that the true value of $\Vrm_{\rm up}$ will be lower in practice because the classical flow will also pull the inflaton down its potential. Below we will include this effect in the convolution integral.


When investigating the effect of the classical flow down the potential along with the two quantum effects, we need to consider the convoluted volume for decreasing values of $N$, from the classical onset of eternal inflation, and down towards Planckian values $N \approx 1$. The magnitudes of the three effects in this regime can be written as:
\begin{subequations}
\begin{align}
	|\dot{N}_{\rm qf}|
		&\simeq
								\frac{n}{2\pi}\.\sqrt[n\,]{\frac{\lambda_{n}}{ 3\.n! }}\,
								N^{\frac{ 1}{ n }}
								\; ,
								\label{eq:NdotofNalpha}
								\displaybreak[1]
								\\[0.5mm]
	|\dot{N}_{\rm dep}|
		&\simeq
								\sqrt{\frac{3}{n | n - 1 |}\,}\,\sqrt[n\,]{ \frac{3\.n!}{\lambda_{n}}}\,
								N^{- \frac{ 2 + n }{ 2n }}
								\; ,
								\displaybreak[1]
								\\[0mm]
	|\dot{N}_{\rm cl}|
		&\simeq
								\left( \frac{ V' }{ V } \right)^{\!2}\!H\,
								N
		\simeq
								n^{2}\.\left(\frac{\lambda_{n}}{ 3\.n! }\right)^{\frac{2}{n}}\,
								N^{\frac{ 4 + n }{ 2n }}
								\; .
\end{align}
\end{subequations}
The classical effect is always positive, whereas the depletion effect is always negative. The quantum fluctuations can take either positive or negative values, but the interesting ones that may lead to eternal inflation are the negative ones. However, as discussed above, this does not mean that the depletion and the quantum fluctuations pull together towards an eternally inflating state. The quantum-depletion effect on the contrary drives the entire state away from its semiclassical description taking the gravitons out of their coeherent state, which makes inflation (eternal or not) meaningless. Thus it is in fact classical flow and the depletion that together both pull the physical state away from the eternally inflating state, but in radically different ways, one by flowing towards the bottom of the potential, and one by destroying the coherence of the underlaying quantum state.

The na{\"\i}ve absolute maximum value for $\lambda_{n}$, which is remotely sensible to consider, is the value for which eternal inflation can begin only at Planck scales $N \simeq 1$ in Eq.~\eqref{eq:N<1/m}, that is $\lambda_{n,\,{\rm max}} \simeq 3\.n!\.(3.8 n)^{-n}$. In practice $\lambda_{n}$ would be much smaller than this as terms of higher order in $1 / N$ become important when $N$ approaches $1$. The minimum value of $\lambda_{n}$ is harder to obtain. However, as $\lambda_n$ decreases, the width of the Gaussian function for the quantum fluctuations decreases, making the quantum-depletion effects more dominant.

For a given value of $n$ and the self-coupling $\lambda_{n}$ we can calculate the fraction of the Universe undergoing eternal inflation by evaluating the convolution integral for a given value of $N$, which is formulated in such a way as to account for the classical flow and the depletion both driving the state away from the potentially eternally inflating state,
\begin{align}
	\Vrm_{\mathrm{up}}
		&\simeq
								\int_{0}^{\infty}\d t\;
								\frac{ \lambda^{t}\,e^{-\lambda} }{ t! }
								\left[
									\int_{- \infty}^{- ( t + t_{\rm cl} )}\d x\;
									\frac{ 1 }{ \sigma\sqrt{2\pi} } e^{- \frac{ x^{2} }{ 2 \sigma^{2} }}
								\right]
								,
								\label{eq:VupGeneral}
\end{align}
with $\lambda = \dot{N}_{\rm dep}$, $\sigma = \dot{N}_{\rm qf}$ and $t_{\rm cl} = \dot{N}_{\rm cl}$ as given in Eqs.~(\ref{eq:NdotofNalpha}-c). The maximum $N$ which corresponds to standard/non-corpuscular eternal inflation reads
\begin{align}
	N_{\rm max}
		&\simeq
								\left(
									\frac{ 3\.n! }{ \lambda_{n} }
								\right)^{\!\! \frac{ 2 }{ n + 2 }}
								\left(								
									\frac{ 1}{ 3.8 n}
								\right)^{\!\!\frac{ 2 n }{ n + 2 }}
								\; .
								\label{eq:N-standard-eternal-inflation}
\end{align}
For instance, $\lambda_{n} = 10^{-12}$ yields $N_{\rm max}( n = 2 ) \sim \Ocal( 10^{6} )$, $N_{\rm max}( n = 3 ) \sim \Ocal( 10^{5} )$, and $N_{\rm max}( n = 4 ) \sim \Ocal( 10^{4} )$. As long as none of the resulting fractions \eqref{eq:VupGeneral} exceed $0.05$, eternal inflation will not occur.

\begin{figure}
	\centering
	\includegraphics[scale=0.46,angle=0]{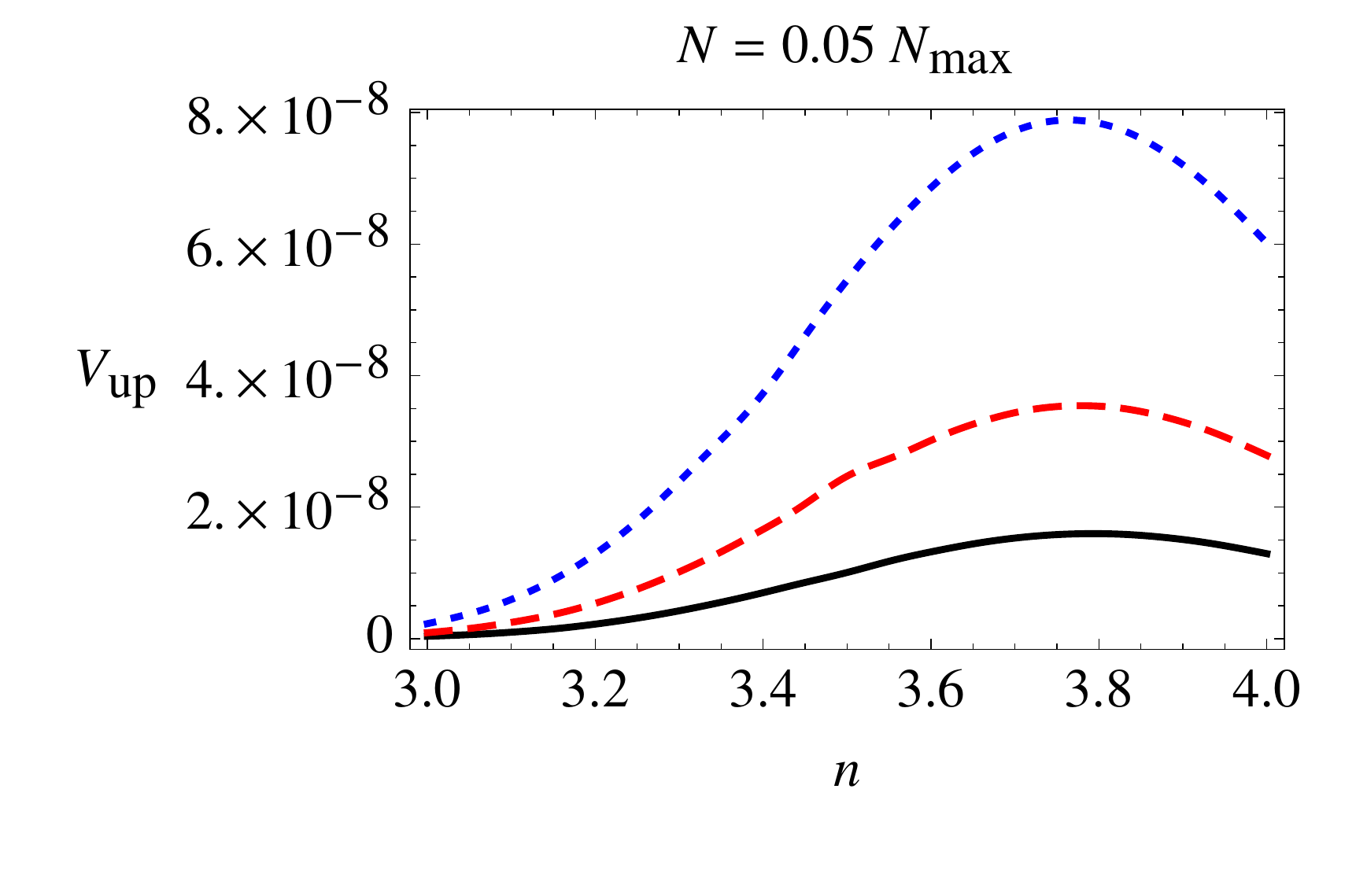}\hspace{-3mm}
	\includegraphics[scale=0.46,angle=0]{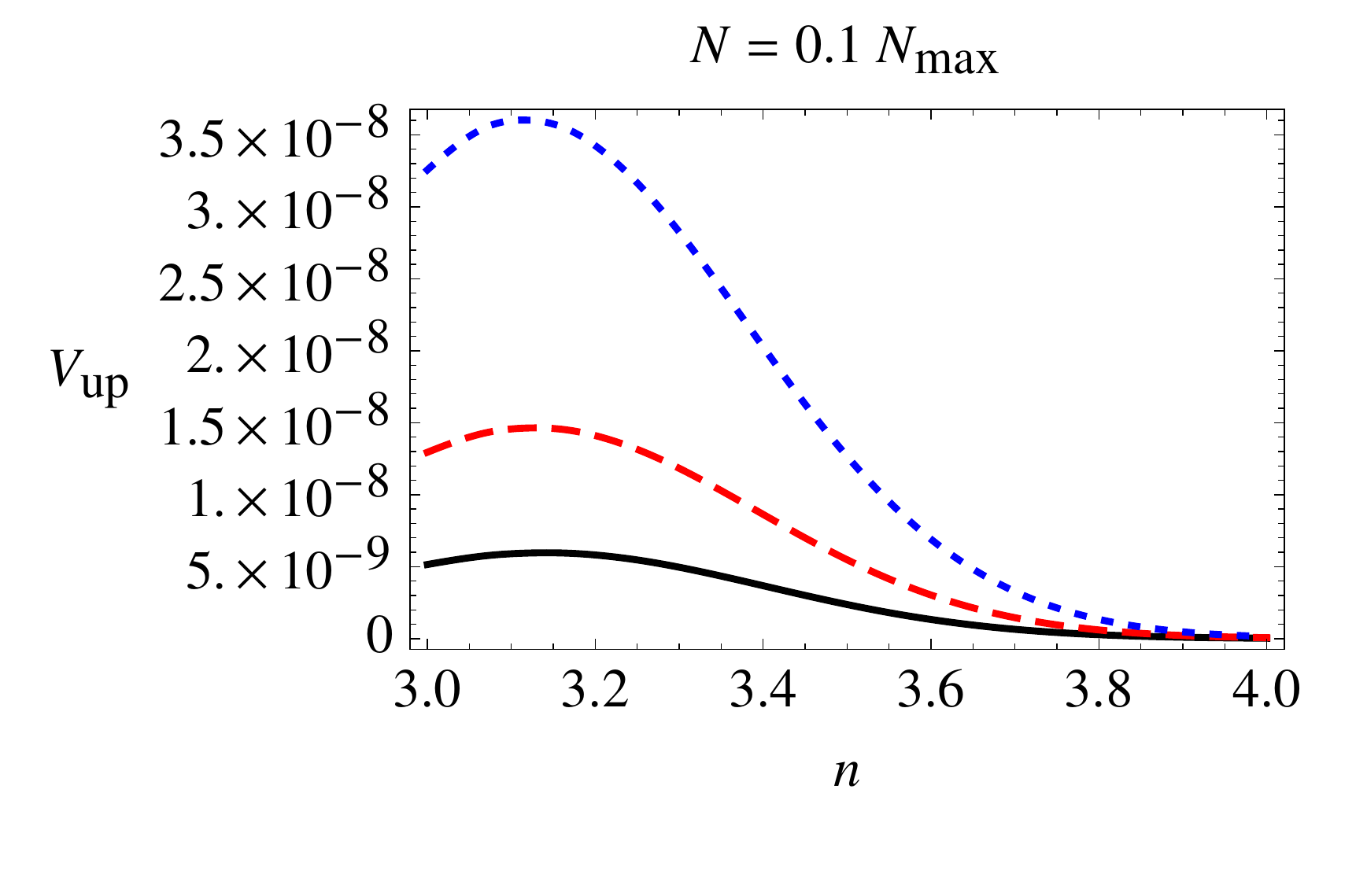}
	\caption{The general convolution integral $\Vrm_{\mathrm{up}}$
			including the classical drift [\cf~Eq.~\eqref{eq:VupGeneral}]
			as a function of the exponent $n$ in $\Vrm( \phi ) = 1 / n!\.\lambda_{n}\.\phi^{n}$
			for various values of the self-coupling $\lambda_{n}$:
			$10^{-8}$ (blue, dotted), $10^{-10}$ (red, dashed), $10^{-12}$ (black, solid);
			{\it Upper panel\.}: $N = 0.05\.N_{\rm max}$; {\it Lower panel\.}: $N = 0.1\.N_{\rm max}$.\\[-4mm]}
	\label{fig:Vup-general}
\end{figure}

In Fig.~\ref{fig:Vup-general} we depict results of the general convolution integral \eqref{eq:VupGeneral} for various values of the self-coupling $\lambda_{n}$ ($10^{-12}$, $10^{-10}$, $10^{-8}$) as well as for two values of $N${\,---\,}once for a tenth of the maximum allowed value as given by \eqref{eq:N-standard-eternal-inflation}, and once for a twentieth of it. For the values considered eternal inflation is dwarfed out by many orders of magnitude.  As $n$ or $\lambda_n$ increases, $N_{\rm max}$ decreases, so for values much larger than the ones considered here, either for $n$ or $\lambda_n$, even considering eternal inflation becomes nonsensical. Hence {\it eternal inflation does not occur for canonical single-field inflation models with monomial self-interactions}.

\section{Discussion and Outlook}
\label{sec:Discussion-and-Outlook}
\setcounter{equation}{0}

\noindent
In this work we investigated the paradigm of eternal inflation in view of the corpuscular picture of space-time for single-field inflation models with monomial potentials. We have compared the strength of the average fluctuation of the field up its potential with that of quantum depletion, and showed that the latter is dominant at least for small $n$.

In order to make a more refined statement, we then studied the fraction of space-time which has an increasing potential both with and without the effects of the classical roll present. For the case where we only considered quantum fluctuations versus quantum depletion, we could already prove the non-existence of eternal inflation for the observationally-relevant small-$n$ potential. 

Including the classical effects we could show that the fraction of space-time moving up the potential is always, \ie~for any $n$, way below the eternal-inflation threshold. Summarized, we have proven that {\it eternal inflation does strictly not occur for all canonical single-field inflation models with monomial self-interactions}. This is a quantitative substantiation of the claim made in \cite{Dvali:2013eja, Dvali:2014gua} that corpuscular gravity prohibits eternal inflation.

We believe that these findings are rather generic. In the case of more general potentials, such as for instance hilltop inflation, we still need the quantum fluctuations to be comparable to the classical evolution in order to drive eternal inflation. In these situations we also expect the depletion effects to become large, and more importantly dominate with respect to the usual quantum fluctuations, more or less regardless of the detailed shape of the potential. Also for many non-monomial potentials, the shape of the potential as seen in the case of an inflaton high enough up in the potential for quantum fluctuations to be comparable to classical flow may be well approximated by a monomial potential.

We should stress that, in any case, at some finite point in time the quantum-depletion effects will accumulate to an extent that invalidates any (semi-)classical treatment, and hence constitutes a radical shift away from the "standard" non-corpuscular description of eternal inflation. Then, the mean-field description is completely different from classical General Relativity and it will be impossible to reliably say that eternal inflation occurs. These statements are completely generic for any corpuscular treatment of inflation, and, in a forthcoming publication, we will further elaborate on this (non-eternal) inflation for generic single-field inflation models, using the quantitative methods developed in this work.

Note that the mentioned physical mechanism investigated in this work 
is very different from all the (semi-)classical ones discussed previously in the literature, which investigate bounds on the total number of e-foldings, originating from extra-dimensions (\cf~\eg~\cite{Wang:2003qr}), by assigning finite entropy to de Sitter-space \cite{Albrecht:2002xs, Banks:2003pt, ArkaniHamed:2007ky}, or, via thoroughly incorporating the null-energy condition \cite{Winitzki:2010yz}, for instance. Instead, here, the limited duration of inflation originates from the quantum resolution of the inflation- as well as of the graviton condensate, which constitute the classical backgrounds in the limit of infinite $N$. In this limit no quantum depletion is present, which, as we quantified, turns out to be crucial for properly deriving the criterion for the occurence of eternal inflation.

As the de Sitter solution might be approximated by the extreme slow-roll version of inflation, the fact that inflationary theories in the near-Planckian range are strongly dominated by the depletion effect also strengthens the argument found in \cite{Dvali:2014gua} that the corpuscular view of gravity may have bearing on the cosmological constant problem.

\acknowledgements
F.~K.~acknowledges supported from the Swedish Research Council (VR) through the Oskar Klein Centre, and thanks the Institute of Theoretical Astrophysics at the University of Oslo where part of this work as been performed. M.S.~thanks David F.~Mota for motivation to discover and explore the field of corpuscular gravity, and is also grateful to Nordita for hospitality during the "Extended theories of Gravity"-workshop. We are indebted to Gia Dvali and Douglas Spolyar for valuable comments.\\[-8mm]


\end{document}